\documentclass[%
 twocolumn,
 reprint,
 showpacs,
 unsortedaddress,
 amsmath,amssymb,
prl,
]{revtex4-1}

\usepackage{graphicx}
\usepackage{bm}

\newcommand{\vect}[1]{{\mbox{\boldmath $#1$}}}
\newcommand{\Eq}[1]{\eqref{eq:#1}}

\newcommand{\Figure}[1]{Figure \ref{fig:#1}}
\newcommand{\Fig}[1]{Fig.\ \ref{fig:#1}}

\newcommand{\deffig}[4]{
\begin{figure}[tb]
  \begin{center}
  \includegraphics[width=#3 \textwidth]{#2}
  \caption{ \label{fig:#1} #4}
  \end{center}
\end{figure}
}

\begin{document}

\preprint{XXXX}

\title{A Scaling Relation for Dangerously Irrelevant Symmetry-Breaking Fields}

\author{Tsuyoshi Okubo}
\email{t-okubo@issp.u-tokyo.ac.jp}
\author{Kosei Oshikawa}
\author{Hiroshi Watanabe}
\author{Naoki Kawashima}
\affiliation{Institute for Solid State Physics, University of Tokyo, Kashiwa 5-1-5, Kashiwa, Japan 277-8581}

\date{\today}

\begin{abstract}
We propose a scaling relation for critical phenomena in which 
a symmetry-breaking field is dengerously irrelevant.
We confirm its validity on the 6-state clock model in three and four dimensions
by numerical simulation.
In doing so, we point out the problem in the previously-used order parameter,
and present an alternative evidence based on the mass-dependent fluctuation.
\end{abstract}

\maketitle



Irrelevant scaling fields are ubiquitous.
While they play minor roles in most cases,
some of them are quite relevant in the usual sense of the word.
A text-book example is the $\phi^4$ term in the
$\phi^4$ theory above the upper critical dimension \cite{CardyText}.
In the present Letter, we discuss cases where such 
a dangerously-irrelevant scaling field reduces the symmetry of the system, 
and demonstrate that it yields a new scaling relation.

Consider a renormalization-group flow diagram including
two fixed points; one describing the critical point and the other the ordered phase.
In principle it is possible that some irrelevant perturbative field
at the critical fixed point contains some scaling field 
that is relevant at the one of the two.
In particular, when the perturbation is symmetry-reducing, 
it can happen that both fixed points lie on the same manifold
characterized by zero of the perturbative field as illustrated in \Fig{RGFlowDiagram}.
In such cases, even if the perturbation almost dies out at some length scale, 
say $\xi$, it may recover its amplitude at larger length scale, say $\xi'$.  
When the system size is between the two scaling
lengths, $\xi \ll L \ll \xi'$, the system may look ordered but still no
effect of the symmetry breaking is visible.  
It may then appear that an intermediate phase exists 
where the system acquires an emergent symmetry.
A classical example of
this type of renormalization group flow is the $q$-state
clock model in three dimensions \cite{Oshikawa2000},
and its continuous-spin counterpart.

In fact, such an intermediate phase really exists in two dimensions \cite{JoseKKN1977}.
However, based on the Monte Carlo simulation results,
Miyashita \cite{Miyashita1997} suggested a simpler scenario for the
three dimensional case.
Furthermore, Oshikawa \cite{Oshikawa2000} pointed out that 
the existence of the intermediate phase is very unlikely
because the low-temperature phase is already ordered in the pure model
in three dimensions, and that the whole low-temperature phase is controlled by
the zero-temperature fixed point, in contrast to the two-dimensional case.
The two-dimensional quantum SU($N$) Heisenberg model may offer 
a quantum-mechanical example.
While the ground state of this model is the Ne\`el state upto $N=4$, 
the valence bond solid state takes over for $N\ge5$ \cite{KawashimaT2007}.
When described in terms of effective spins representing the direction
of the ordered valence bond pattern, the system can be regarded as
a model analogous to the clock model.
It was discovered that the order parameter distribution function is 
almost circular symmetric, 
indicating the extremely small effect of the anisotropy.  
Later, an additional term was 
introduced \cite{Sandvik2007,HaradaKT2007,HaradaSOMLWTK2013} 
to control the quantum fluctuation and drive the system 
to the true transition point.


\deffig{RGFlowDiagram}{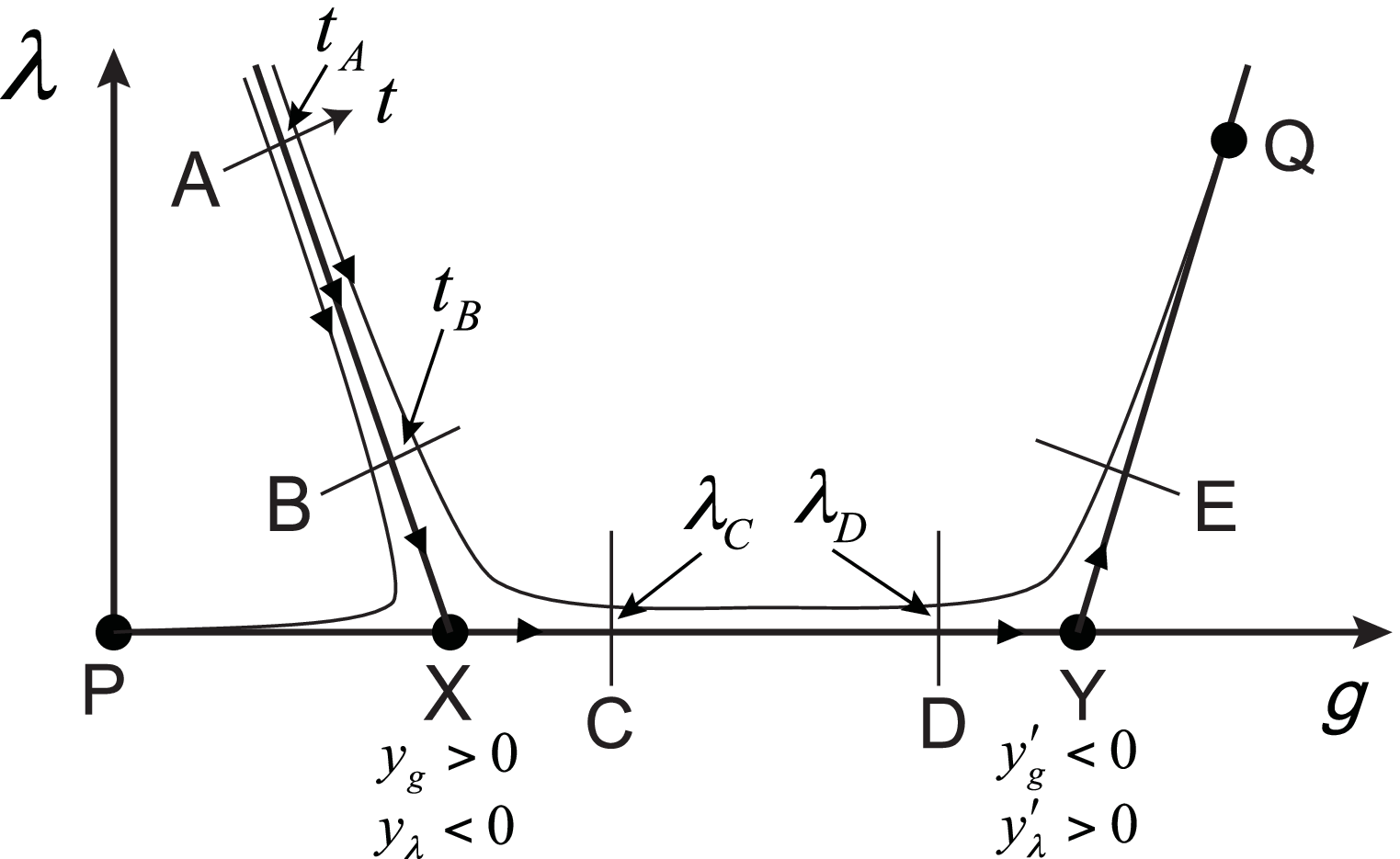}{0.42}{ The generic
  renormarization flow diagram with four fixed points: P, Q, X, and Y.
  }

It is now widely accepted that in three dimensions there is no
partially ordered phase with the emergent symmetry. 
However, disagreement still persists concerning the scaling relation
that relates the scaling exponent $\nu'$ that characterizes the
longer correlation length and $\nu$ characterizing the shorter 
correlation length.
In this Letter, we propose a new general scaling relation and 
verify its validity by Monte Carlo simulation of the
$XY$ model with the $Z_q$ scaling field. To verify the validity of the new
scaling relation, below we first present the numerical results of the
anisotropy order parameter, often referred to as $\phi_6$, suggesting
that previously-proposed scaling relations do not actually hold. 
We further argue that, unlike the conventional finite-size scaling, the scaling
plot of $\phi_6$ is not fully supported by renormalization group picture;
we present a more complete scaling argument supported by Monte Carlo
simulation.


Previously, a scaling relation was proposed by Ueno {\it et al.} \cite{UenoM1991}
and by Oshikawa \cite{Oshikawa2000}.
Their argument is based on the basic assumption that there is a well
defined domain wall splitting the whole system and the excess free-energy caused by the domain walls 
is the scaling variable.
The excess free-energy density per area of the domain wall may be 
given by the symmetry-breaking field renormalized upto the 
scale of locally-correlated volume, $\lambda(\xi) \sim \lambda
\xi^{y_{\lambda}}$ ($y_{\lambda}$ represents the scaling exponent of the
symmetry-breaking field at the critical fixed point).
The total domain-wall free-energy, then, may be
$L^{d-1} \lambda(\xi) \sim (L/\xi^{-y_{\lambda}/(d-1)})^{d-1}$.
This yields
\begin{equation}
  \frac{\nu'}{\nu} = \frac{-y_{\lambda}}{d-1}.
  \label{eq:ScalingRelationI}
\end{equation}
Lou, Sandvik, and Balents \cite{LouSB2007} presented a similar argument,
but they argued that the effect of the anisotropy free-energy comes from
the volume instead of the domain-walls.
Therefore, they multiply the renormalized field by the number of
correlated volumes, to obtain $(L/\xi)^d \lambda \xi^{y_{\lambda}}$
$= \lambda (L/\xi^{1-y_{\lambda}/d})^d$. This means
\begin{equation}
  \frac{\nu'}{\nu} = 1 + \frac{-y_{\lambda}}{d}.
  \label{eq:ScalingRelationII}
\end{equation}


Here we present another scaling relation that is more general
and differs from the previous ones.
We again consider the generic renormalization group flow of \Fig{RGFlowDiagram}.
The bare Hamiltonian is along the short line near the point ``A''
parametetrized by $t$ so that $t=0$ corresponds to the critical point.
If we start from the point $t=0$ on this line,
the scaling flow takes us to the critical fixed point ``X'', where
$g = g_{\rm X}$ and $\lambda = 0$.
If we start from a point with $t=t_{\rm A}>0$ and $\lambda=\lambda_{\rm
A} > 0$, 
the scaling flow goes through the points
``C'' $(|g_{\rm C} - g_{\rm X}| = O(1)$, $\lambda_{\rm C} \ll 1)$,
``D'' $(|g_{\rm D} - g_{\rm Y}| = O(1)$, $\lambda_{\rm D} \ll 1)$,
and approaches the second fixed point ``Y''
around which renormalization group flow is characterized by scaling exponents $y'_g < 0$ for the variable $g$
and $y'_{\lambda} > 0$ for the variable $\lambda$.
Because of the presence of $\lambda$, the flow deviates from ``Y'',
goes through the point ``E'' $(\lambda_{\rm E} = O(1))$
and eventually reaches some other fixed point.
The shorter correlation length $\xi$ equals $\Lambda_{\rm AC}$,
i.e., the length scale that has to be renormalized to go from ``A'' to ``C'', 
whereas the longer correlation length $\xi'$ equals $\Lambda_{\rm AE}$.
The critical intervals are $BC$ and $DE$. 
For the interval $BC$, we have
$
  \lambda_{\rm C} \sim \xi^{y_{\lambda}}.
$
For $DE$, 
$
  1 \sim \lambda_{\rm E} \sim \lambda_{\rm D} (\Lambda_{\rm DE})^{y'_{\lambda}},
$
which yields
$
  \Lambda_{\rm DE} \sim 
  (\lambda_{\rm D})^{-1/y'_{\lambda}}
$
$
  \propto 
  (\lambda_{\rm C})^{-1/y'_{\lambda}}
$
$ 
  \sim \xi^{-y_{\lambda}/y'_{\lambda}}.
$
Therefore,
$$
  \xi' \propto \Lambda_{\rm BC} \Lambda_{\rm DE}
  \sim \xi^{1 + \frac{-y_{\lambda}}{y'_{\lambda}}}.
$$
Thus we have arrived at
\begin{equation}
  \frac{\nu'}{\nu} = 1 + \frac{-y_{\lambda}}{y'_{\lambda}}.
  \label{eq:ScalingRelationIII}
\end{equation}

%
In order to determine which scaling relation should apply,
we need independent estimates of the scaling indices,
$\nu$, $\nu'$, $y_{\lambda}$, and $y'_{\lambda}$, in \Eq{ScalingRelationIII}.
Here we consider the $XY$ model in three dimensions with the $Z_q$ anisotropy field.
$$
  H = -J \sum_{(\vect{r},\vect{r}')} \cos(\theta(\vect{r}) - \theta(\vect{r}'))
  -\lambda_q \sum_{\vect{r}} \cos(q\theta(\vect{r}))
$$
As for $\nu$, previous estimates of the pure $XY$ universality class is available,
$\nu = 0.6717(1)$ \cite{CampostriniHPV2006}.
As for $y_{\lambda}$, previous calculation according to the
first-order $\varepsilon$-expansion \cite{Oshikawa2000} leads,
$$
  y_{\lambda} = 4 - q + \varepsilon\left( 
  \frac{q}{2} - 1 - \frac{q(q-1)}{10}
  \right),
$$
e.g., $y_{\lambda} = -0.2$ for $q=4$ and $=-3.0$ for $q=6$.
In addition to this $\varepsilon$-expansion, Monte Carlo estimates of the
$y_{\lambda}$ upto $q=4$ are available \cite{HasenbuschV2011}.
In \Fig{yq_extrapolation}, we plot the estimated scaling eigenvalues 
and their extrapolation by the second-order polynomial along with the
result of the first-order $\varepsilon$-expansion. The Monte Carlo
estimation of $y_{\lambda}$s reveals a surprisingly good agreement with the
first-order $\varepsilon$-expansion, while the second-order polynomial
fitting is slightly deviate from the $\varepsilon$-expansion at $q=6$.
From this figure, we estimate $y_{\lambda} = -2.5(2)$ for $q=6$. 
\deffig{yq_extrapolation}{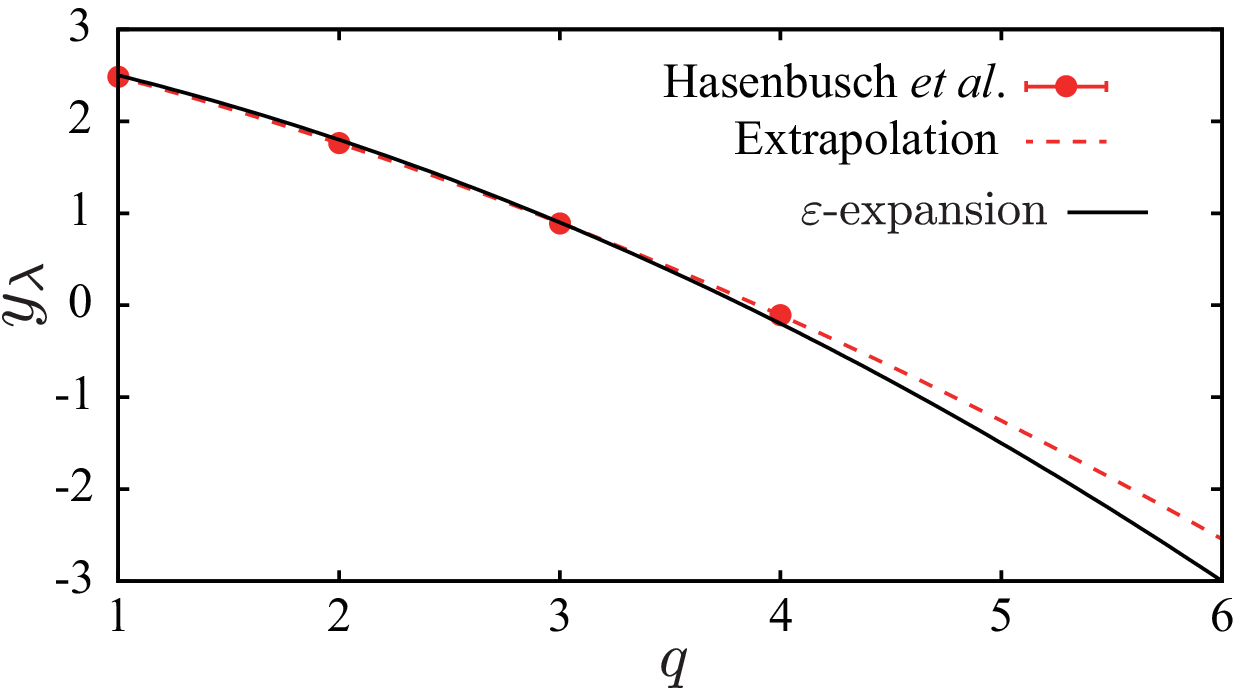}{0.4}{(color online)
  The estimated scaling eigenvalues from Ref.~\cite{HasenbuschV2011}(symbols), 
  the second-order polynomial fitting to them (dashed curve), 
  and the result of the first-order $\varepsilon$-expansion (solid curve).
}
%
As for $y'_{\lambda}$, an argument \cite{Oshikawa2000} suggests 
that the quadratic fluctuation around the ordered configuration 
is essential at the NG fixed point,
leading to $y'_{\lambda} = 2$, analogous to the scaling eigenvalue 
of the $\phi^2$ field in the Gaussian field theory. 
Finally, we consider $\nu'$. In order to estimate $\nu'$ we need a
proper scaling variable which obeys a finite size scaling with
$\nu'$. 

In the previous studies, an order parameter that characterizes the
symmetry reduction from $U(1)$ to $Z_6$, 
$$
\phi_6 \equiv \langle \cos(6\theta_0) \rangle
$$ 
was analyzed by assuming $\phi_6 \sim
f(tL^{1/\nu'})$ \cite{Oshikawa2000,LouSB2007}.
Here, $\theta_0$ is the angle of the average magnetization, i.e., 
$$
  (m_0 \cos\theta_0, m_0\sin\theta_0) \equiv 
  \frac1{N} \sum_{\vect{r}} (\cos\theta(\vect{r}),\sin\theta(\vect{r})),
$$
and $\langle \cdots \rangle$ represents a thermal average.
In \Fig{phi6}(a), we show the finite size scaling of $\phi_6$ against
$(T_c - T)L^{1/\nu'}$ with $\nu' = 1.45$ which is estimated from
Bayesian method \cite{Harada:2011js}. Estimated $\nu'$ considerably
deviated from Ueno's scaling relation \Eq{ScalingRelationI} and Lou's
scaling relation \Eq{ScalingRelationII}; they give $\nu' \simeq 0.84$ and
$\nu' \simeq 1.23$ from known exponents $\nu$ and $y_{\lambda}$,
respectively. Indeed, when we use even relatively acceptable Lou's
value, $\nu'=1.23$, the overlap of the data become clearly worse (see
\Fig{phi6}(b)). Therefore scaling relations \Eq{ScalingRelationI} and
\Eq{ScalingRelationII} seem to fail in the present critical phenomenon.

\deffig{phi6}{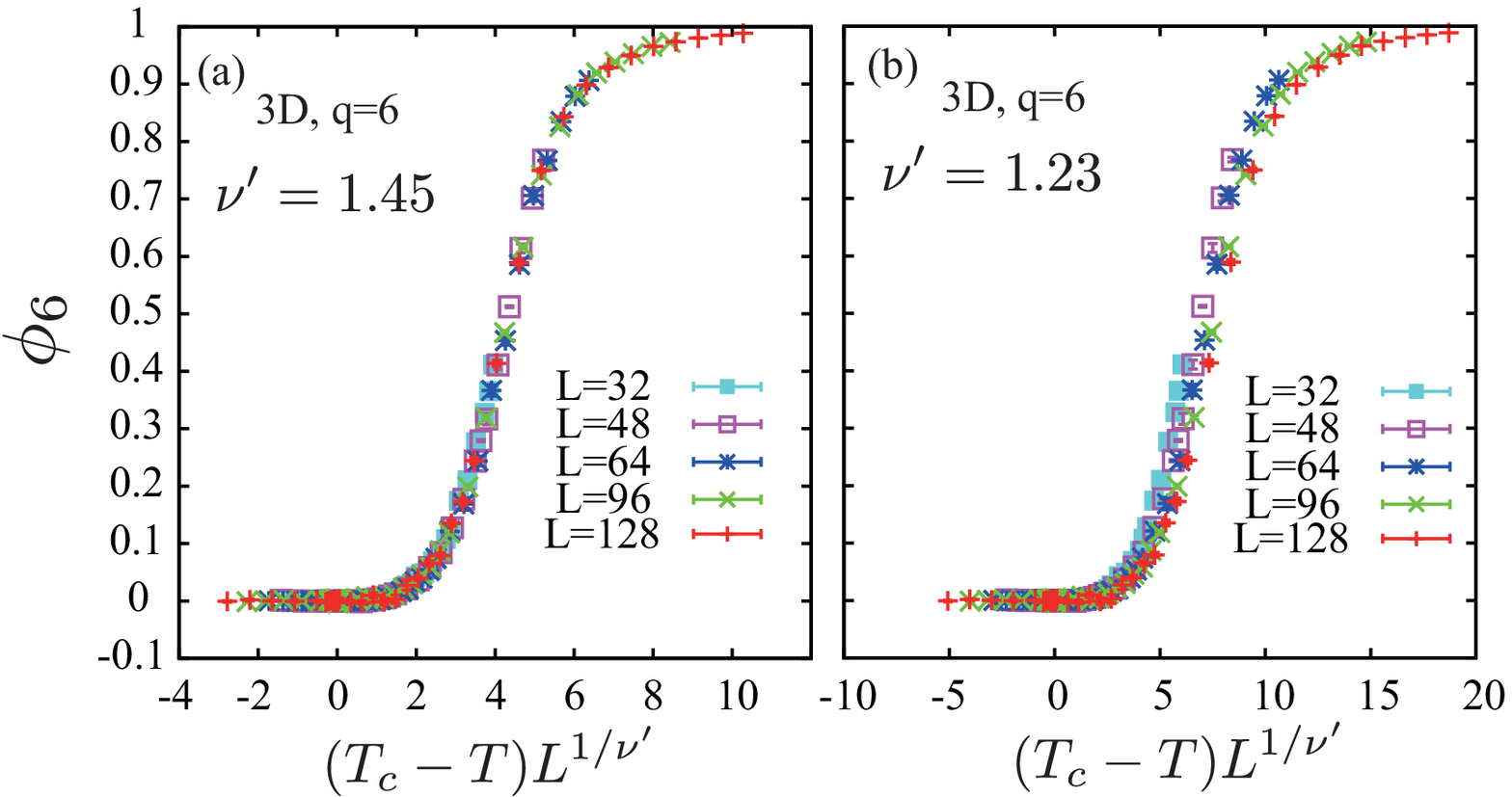}{0.47}{(color online) The finite size
  scaling of $\phi_6$ for the $Z_6$-anisotropic $XY$ model with
  $\lambda_6/J = 5$ in three dimensions. (a) $\nu' = 1.45$, (b) $\nu'
  =1.23 $. For the scaling we used the date set of $(T_c - T)/J \le 0.4$.}

Note, however, this finite-size scaling plot of $\phi_6$ is different
from conventional finite-size scaling in that it is not clear whether
we would obtain a perfect data collapse even in the limit of infinite system size.
More specifically, to regard $\phi_6$ as a scaling operator in
its full value range, $\theta_0$ must be dimensionless scaling
operator. However, at the NG fixed point, it has a non-zero scaling
dimension \cite{Oshikawa2000}. Therefore, the meaning of scaling
analysis based on $\phi_6$ is not clear, even if the resulting plottings
may look reasonably good \footnote{The reason why \Fig{phi6} shows good
``data collapse'' may be that $\phi_6$ becomes finite only when the
renormalized annisotropy parameter $\tilde\lambda_q \equiv \lambda'_q
(L/\xi)^2$ is finite. Since the latter is essentially the $x$-variable
of \Fig{phi6}, all the curves depart from the $x$-axis more or less at
the same value of $x$.}. We propose another scaling variable
whose scaling form is directly calculated from the effective theory
around the NG fixed point.

Suppose that we start from the high-temperature phase and gradually cool
the system passing the transition point. Because of the asymptotic U(1)
symmetry, the ordering angle $\theta_0$, selected by the spontaneous
symmetry-breaking, can be any value in the interval $[0,2\pi)$. Once the
ordering angle has been selected, it does not change (in a finite time)
and determines the ``mass of the particles''. More specifically, the
effective Hamiltonian that characterizes the system at the length scale
larger than the (first) correlation length $\xi$ can be obtained by
expanding
$ \mathcal{H} = \int d^d\bm{r}' \left[\frac{1}{2}(\bm{\nabla}\theta)^2 - \lambda_q' \cos q \theta\right]
$
in terms of the small fluctuation around the direction of the spontaneous ordering,  $\phi \equiv \theta - \theta_0$:
\begin{multline}
 \mathcal{H}_{\theta_0} = \frac{1}{2}\int d^d\bm{r}' \left[(\bm{\nabla}\phi)^2 + \lambda_q' q^2 \cos (q
		     \theta_0) \phi^2 \right]\\
 - \lambda_q' L'^d
 \cos q\theta_0 + O(\phi^3).
\end{multline}
Note that $\bm{r}'$ implies the renormalized length $\bm{r}' \sim \bm{r}/\xi $.
This Gaussian field theory indicates that Fourier modes of fluctuation
are governed by renormalized anisotropy and macroscopic orientation
$\theta_0$ as 
\begin{equation}
\langle|\phi_{\bm{k}'}|^2\rangle_{\theta_0} \sim \frac{1}{k'^2 +
 \lambda'_q q^2\cos q \theta_0},
  \label{eq:phi_fluctuation}
\end{equation} 
where $\langle \cdots \rangle_{\theta_0}$ means the average with the
condition that the macroscopic orientation is equal to $\theta_0$.
Note that every quantity in this expression is normalized upto the length scale $\xi$, i.e.,
$\lambda'_q = \lambda_q \xi^{y_{\lambda}}$, $\vect{r}' \equiv \vect{r}/\xi$, and
$k' \equiv k \xi$. Then, if we take the wavenumber $k'$
as $2\pi/(L/\xi)$,
$\langle|\phi_{\bm{k}'}|^2\rangle_{\theta_0}$ obeys the scaling form $\langle|\phi_{\bm{k}'}|^2\rangle_{\theta_0} \sim (L/\xi)^2 f(L/\xi^a)$ with $a \equiv 1 - y_{\lambda}/2$. This form indicates that $\xi' \sim \xi^a$ or $\nu'/\nu = a$. 


\deffig{Histogram}{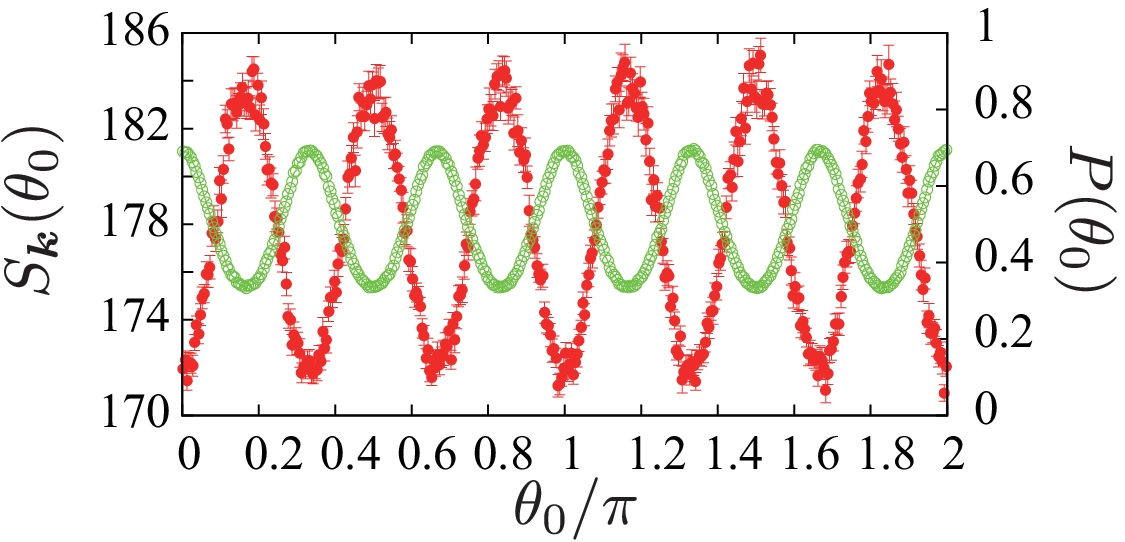}{0.4}{(color online)
  The angle dependent spin structure factor $S_{\bm{k}}(\theta_0)$ (filled
  circle) and the probability distribution of ordering angle $\theta_0$
  (open circle) for the $Z_6$-anisotropic $XY$ model with
  $\lambda_6/J = 5$, $L=64$ at $T/J = 2.0 < T_c/J$ ($T_c/J = 2.202$) in three dimensions.}

\deffig{StructureFactor}{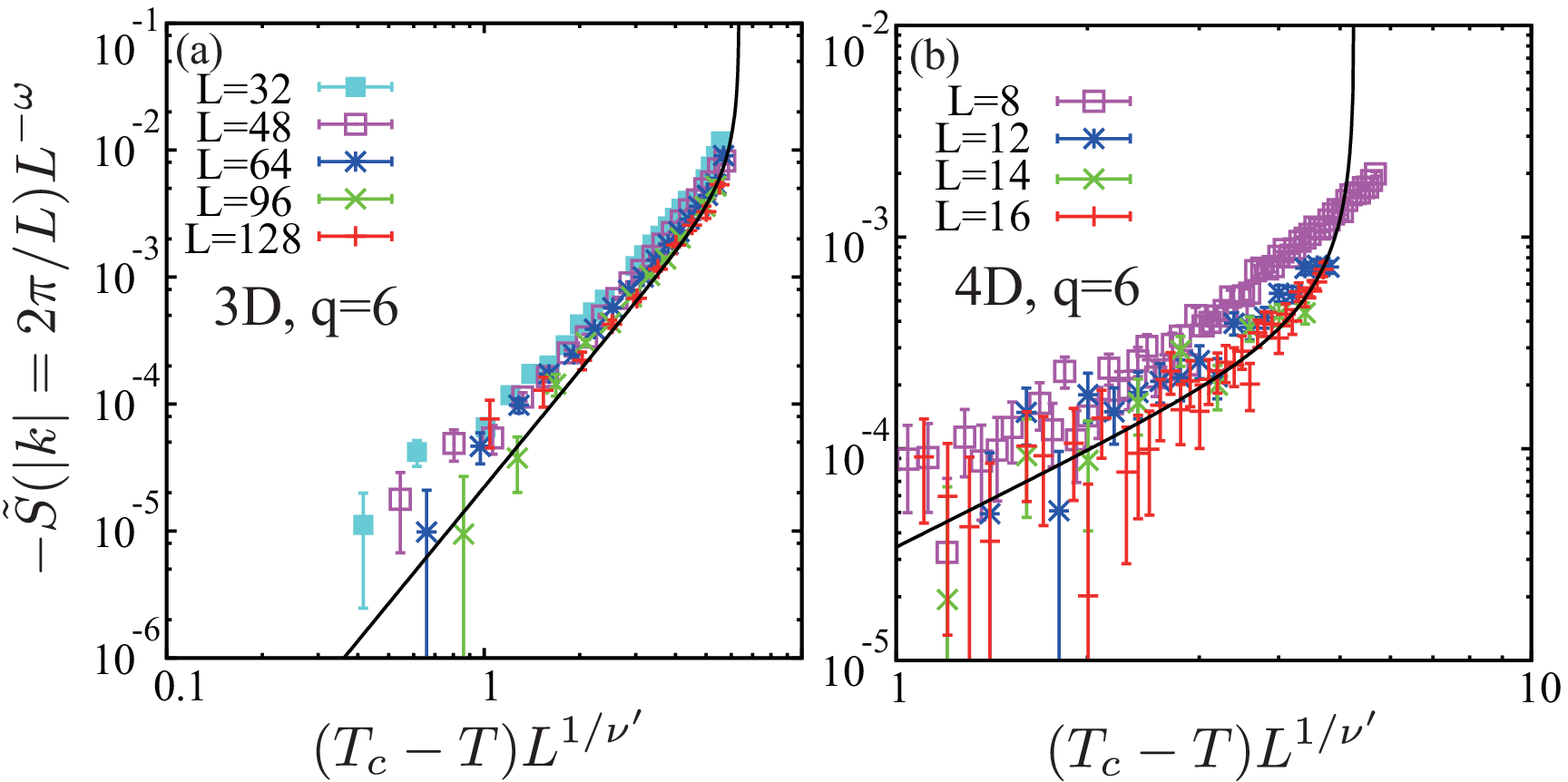}{0.47}{(color online) Log-log plot of the angular Fourier transform
  of the angle dependent spin-structure factor $\tilde{S}$ for the
  $Z_6$-anisotropic $XY$ model with $\lambda_6/J = 5$ in three
  dimensions (a) and in four dimensions(b).  (a) the
  finite-size scaling plot assuming $\omega = 1.983 $ and $\nu' = 1.511$
  . (b) the finite-size scaling plot assuming $\omega = 2.0 $ and
  $\nu' = 1$ in four dimension. The critical temperatures $T_c/J =
  2.202$ for three dimension and $T_c/J = 3.3143$ for four dimenstions are estimated from the
  binder ratios of the magnetization. The solid curves are the fittings of
  the scaling function \Eq{ScalingFunction} assuming
  $\nu = 0.6717$, $x_h = 0.519$, $y_{\lambda} =-2.5$ for three
  dimensions and $\nu = 0.5$, $x_h = 1$, $y_{\lambda} =-2$ for four
  dimensions with the relation $\nu'/\nu = 1 - y_{\lambda}/2$.}

In order to estimate this exponent, we
carried out Monte Carlo simulation of the $XY$ model with anisotropy,
$\lambda_6 = 5J$. We computed the spin structure factor at $|\bm{k}| =
2\pi/L$ as an observable for the finite size scaling. Based on the
effective Hamiltonian around the NG fixed point, the spin structure factor
is also expected to depend on the ordering direction of the bulk,
$\theta_0$. 
%
%
The angle dependent spin structure factor, 
$
S_{\bm{k}}(\theta_0)\equiv \frac{1}{N}\langle |\sum_{i}\vec{S}_i e^{i\bm{k}\cdot\bm{r}_i}|^2 \rangle_{\theta_0}
$, is naturally related to the Fourier transform of
the renormalized angle fluctuation,
$\langle|\phi_{\bm{k}'}|\rangle_{\theta_0}$, through the relation
\begin{equation}
 S_{\bm{k}}(\theta_0) \sim \xi^{d-2x_h} \langle|\phi_{\bm{k}'}|^2\rangle_{\theta_0},
\label{eq:Sq_phi_Relation}
\end{equation}
where the prefactor $\xi^{d-2x_h}$ with the scaling dimension of the
magnetization at the XY critical fixed point, $x_h$, comes from the
renormalization effect. \Figure{Histogram} shows an example of
$S_{\bm{k}}(\theta_0)$ below the critical temperature $T_c$. As expected
from the behavior of $\langle|\phi_{\bm{k}'}|\rangle_{\theta_0}$,
$S_{\bm{k}}(\theta_0)$ shows a periodic change of its amplitude with the
period of $2\pi/6$. In addition, the minimum appears at the maximum of
the distribution function, which is consistent with
\Eq{phi_fluctuation}.

In order to capture the scaling behaviour of this angle- (or mass-)
dependent fluctuation, we define the angular Fourier transform of the
spin structure factor: $ \tilde{S}_{\bm{k}} \equiv \frac{1}{2\pi}
\int_0^{2\pi} d \theta_0 S_{\bm{k}}(\theta_0) \cos(q \theta_0)$.
From equations \Eq{phi_fluctuation} and \Eq{Sq_phi_Relation}, we expect
that the finite-size scaling of $\tilde{S}_{\bm{k}}$ is given as
\begin{equation}
 \tilde{S}_{\bm{k}} \sim L^{\omega} g((T_c- T)L^{1/\nu'}),
\end{equation}
where
\begin{equation}
 \omega = \frac{2 (d - y_{\lambda} -2 x_h)}{2-y_{\lambda}}.
  \label{eq:OmegaRelation}
\end{equation}
In addition, we can write down the expected scaling function $g(x)$ from the
angular Fourier transform of \Eq{phi_fluctuation} as 
\begin{equation}
 g(x) \propto x^{\nu(2x_h + y_{\lambda} - d)}\left(1 - \frac{1}{\sqrt{1 - c x^{2\nu'}}}\right),
\label{eq:ScalingFunction}
\end{equation}
where $c$ is a non-universal constant.


\Figure{StructureFactor} (a) shows the finite size scaling of
$\tilde{S}_{\bm{k}}$ for the three dimensional ($d=3$) model against
$(T_c-T) L^{1/\nu'}$ with $\nu' = 1.511$ which is calculated from the
known exponents $\nu = 0.6717$ and $y_{\lambda} = -2.5$ through the new
scaling relation \Eq{ScalingRelationIII}. The vertical axis is also
scaled by $L^{-\omega}$ with $\omega = 1.983$ estimated from the
exponents $\nu, y_{\lambda}$ and $x_h =
0.519$(\cite{CampostriniHPV2006}) by \Eq{OmegaRelation}. The data seem
to be almost converged for the larger sizes and also its scaling
function is well fitted by the expected function
\Eq{ScalingFunction}. These observations strongly support the the new
scaling relation \Eq{ScalingRelationIII}.



In retrospect, even if we take the line of reasoning of Ueno {\it et
al.}, we might have had to multiply $(L/\xi)^{d-1}$ instead of $L^{d-1}$
because we are working with the renormalized world with the original
length scale $L$ being shrunk to $L/\xi$.  If we adopt this correction,
Ueno's scaling relation\Eq{ScalingRelationI} would have been $$
\frac{\nu'}{\nu} = 1 + \frac{-y_{\lambda}}{d-1}, $$ which yields an
identical result to the present scaling relation for $d=3$.  
In \Fig{StructureFactor}(b), we plot the finite size scaling
of the $\tilde{S}_{\bm{k}}$ for the four-dimensional ($d=4$) model along
with the fitting curve of the scaling function
\Eq{ScalingFunction}. Although we still observe the strong finite size
correction, the data seem to converge to the scaling function
\Eq{ScalingFunction} supporting the new scaling relation
\Eq{ScalingRelationIII} also in four dimensions.

In summary, we have proposed a generic scaling relation for critical
phenomena in which a dangerously irrelevant scaling field plays an
important roll. 
Monte Carlo simulations for $XY$ model with $Z_q$ symmetry
breaking field strongly supported the validity of new scaling
relation. While we could have used more conventional quantities, such as
the specific heat for instance, to demonstrate the new scaling relation,
we have not done that mainly because of technical difficulty. The effect
of $\xi'$ would appear only as the crossover between two temperature
regimes, one with a sub-dominant contribution from the phase fluctuation
and the other without, offering an evidence much less clear than the one
presented above.

\begin{acknowledgements}
We owe helpful discussions to M.~Oshikawa, H.~Tsunetsugu, K.~Ueda, J.~Lou, 
and S.~Miyashita.
The computation in the present work is executed on computers 
at the Supercomputer Center, ISSP, University of Tokyo. 
The present work is financially supported by MEXT Grant-in-Aid for 
Scientific Research (B)(25287097), and by CMSI, MEXT-SPIRE, Japan.
\end{acknowledgements}

\bibliography{disf2014}

\end{document}